# Discrete Sampling Theorem, Sinc-lets and Other Peculiar Properties of Sampled Signals

Leonid P. Yaroslavsky,
*(Tampere International Center for Signal Processing, Tampere University of Technology, Tampere, Finland)*
*(On leave from Dept. of Physical Electronics, School of Electrical Engineering, Tel Aviv University, Tel Aviv, Israel)*
*e-mail: yaro@eng.tau.ac.il*

**Abstract**

Discrete sampling theorem is formulated that refers to discrete signals specified by a finite number of their samples and band-limited in a domain of a certain orthogonal transform. Conditions of the recoverability of such signals from their sparse samples are discussed for different transforms and applications are illustrated by examples of image super-resolution from multiple chaotically sampled frames and in image reconstruction from projections. Experimental evidence is presented of the existence of discrete signals sharply bounded both in space and DFT or DCT domains and of the family of the corresponding basis functions.

## 1. Introduction

Sampled signals are used in digital signal processing as representatives of continuous signals from which they are obtained by sampling. Sampled signals are specified in computers by a finite number of samples and are believed to sufficiently well approximate original continuous signals by means of an appropriate interpolation of their samples. At the same time sampled signals have a number of peculiar properties that have no exact analogies in continuous signals they represent.

In particular, sampled signals of $N$ samples can be sharply band-limited in a certain transform domain, whereas real continuous signals are never sharply band-limited. For such signals, the paper formulates and proves, the Discrete Sampling Theorem that states that any sampled signal of $N$ samples known to have only $K \le N$ non-zero transform coefficients for certain transform $\Phi_N$ ($\Phi_N$ - transform "band-limited" signal) can be precisely recovered from exactly $K$ of its samples provided positions of the samples secure the existence of the matrix $\mathbf{KofN}_{\Phi}^{-1}$ inverse to the sub-transform matrix $\mathbf{KofN}_{\Phi}$ that corresponds to the band-limitation. For such transforms as DFT and DCT, it is proven that precise recovery of DFT- or DCT-band-limited 1-D signals is always possible with no restrictions to positions of available samples. For multi-dimensional signals, it is shown that this property holds for separable band-limitation. The paper illustrates how this property can be used for precise recovery or MSE-optimal approximation of sampled signals from sparse data. The condition for recoverability of signals band-limited in other transforms, such as Walsh, Haar, other wavelets and Radon Transforms, are shown to be more restrictive in terms of positions of available signal samples. These topics are discussed in Sect. 2 that is based on materials published in Ref. 1

Yet another peculiar property of sampled signals discussed in the paper is the existence of sharply DFT or DCT- band-limited signals with sharply limited support, a property that contradicts the well-known uncertainty principle for continuous signals. In Sect. 3 we present experimental evidence of the existence of such 1-D and 2-D signals and an algorithm for generating band-limited and space limited versions of real signals and formulate, for the entire family of DFT or DCT band-limited and space limited signals, a discrete "uncertainty" relationship between signals and their DFT or DCT spectra, according to which space-bandwidth product $K_S \times K_T$ of the number of non-zero samples of signals and their DFT or DCT spectral coefficients, respectively, is of the order of the total number $N$ of signal samples. On this concept, in Sect 4 a new family of sinc-let basis functions is introduced and experimentally studied that can be used for representation of DFT or DCT- band-limited and space limited signals.

## 2. Discrete sampling theorem and signal recovery from sparse or non-uniformly sampled data

### 2.1 Discrete sampling theorem: formulation

Let $\mathbf{A}_N$ be a vector of $N$ samples $\{a_k\}_{k=0,...,N-1}$, which completely define a discrete signal, $\Phi_N$ be an $N \times N$ orthogonal transform matrix,

$$\Phi_N = \{\varphi_r(k)\}_{r=0,1,...,N-1} \tag{2.1}$$

composed of basis functions $\varphi_r(k)$, and $\Gamma_N$ be a vector of signal transform coefficients $\{\gamma_r\}_{r=0,...,N-1}$ such that:

$$\mathbf{A}_N = \Phi_N \Gamma_N = \left\{ \sum_{r=0}^{N-1} \gamma_r \varphi_r(k) \right\}_{k=0,1,...N-1} \tag{2.2}$$

Assume now that available are only $K < N$ signal samples $\{a_{\tilde{k}}\}_{\tilde{k}\in\tilde{\mathbf{K}}}$, where $\tilde{\mathbf{K}}$ is a $K$-size non-empty subset of indices $\{0,1,..,N-1\}$. These available $K$ signal samples define a system of $K$ equations:

$$\left\{ a_k = \sum_{r=0}^{N-1} \gamma_r \varphi_r(k) \right\}_{k\in\tilde{\mathbf{K}}} \tag{2.3}$$

for signal transform coefficients $\{\gamma_r\}$ of certain $K$ indices $r$.

Select now a subset $\tilde{\mathbf{R}}$ of $K$ transform coefficients indices $\{\tilde{r} \in \tilde{\mathbf{R}}\}$ and define a "$\mathbf{KofN}$"-band-limited approximation $\hat{A}_N^{BL}$ to the signal $\mathbf{A}_N$ as:

$$\hat{\mathbf{A}}_N^{BL} = \left\{ \hat{a}_k = \sum_{\tilde{r}\in R} \gamma_{\tilde{r}} \varphi_{\tilde{r}}(k) \right\} \tag{2.4}$$

Rewrite this equation in a more general form:

$$\hat{\mathbf{A}}_N^{BL} = \left\{ \hat{a}_k = \sum_{r=0}^{N-1} \tilde{\gamma}_r \varphi_r(k) \right\} \tag{2.5}$$

and assume that all transform coefficients with indices $r \notin \tilde{\mathbf{R}}$ are set to zero:

$$\tilde{\gamma}_r = \begin{cases} \gamma_r, & r \in \tilde{R} \\ 0, & otherwise \end{cases} \tag{2.6}$$

Then the vector $\tilde{\mathbf{A}}_K$ of available signal samples $\{a_{\tilde{k}}\}$ can be expressed in terms of the basis functions $\{\varphi_r(k)\}$ of transform $\Phi_N$ as:

$$\tilde{\mathbf{A}}_K = \mathbf{KofN}_\Phi \cdot \tilde{\Gamma}_K = \left\{ a_{\tilde{k}} = \sum_{\tilde{r}\in R} \gamma_{\tilde{r}} \varphi_{\tilde{r}}(\tilde{k}) \right\} \tag{2.7}$$

where $K \times K$ sub-transform matrix $\mathbf{KofN}_\Phi$ is composed of samples $\varphi_{\tilde{r}}(\tilde{k})$ of the basis functions with indices $\{\tilde{r} \in \tilde{\mathbf{R}}\}$ for signal sample indices $\tilde{k} \in \tilde{\mathbf{K}}$, and $\tilde{\Gamma}_K$ is a vector composed of the corresponding sub-set $\{\gamma_{\tilde{r}}\}$ of complete signal transform coefficients $\{\gamma_r\}$. This subset of the coefficients can be found as,

$$\tilde{\Gamma}_K = \{\tilde{\gamma}_r\} = \mathbf{KofN}_\Phi^{-1} \cdot \tilde{\mathbf{A}}_K \tag{2.8}$$

provided matrix $\mathbf{KofN}_\Phi^{-1}$ inverse to the matrix $\mathbf{KofN}_\Phi$ exists, which, in general, is conditioned, for a specific transform, by positions $\tilde{k} \in \tilde{\mathbf{K}}$ of available signal samples and by the selection of the subset $\{\tilde{R}\}$ of transform basis functions.

By virtue of the Parceval's relationship for orthonormal transforms, the band-limited signal $\hat{\mathbf{A}}_N^{BL}$ approximates the complete signal $\mathbf{A}_N$ with mean squared error:

$$MSE = \left\| A_N - \hat{A}_N \right\|^2 = \sum_{k=0}^{N-1} \left| a_k - \hat{a}_k \right|^2 = \sum_{r \notin R} \left| \gamma_r \right|^2 \tag{2.9}$$

This error can be minimized by an appropriate selection of the $K$ basis functions of the sub-transform $\mathbf{KofN}_\Phi$. In order to do so, one must know the energy compaction ordering of basis functions of the transform $\Phi_N$. If, in addition, one knows, for a class of signals, a transform that features the best energy compaction in the smallest number of transform coefficients, one can, by selection of this transform, secure the best band-limited approximation of the signal $\{a_k\}$ for the given subset $\{\tilde{a}_k\}$ of its samples.

In this way, we arrive at the following Discrete Sampling Theorem that can be formulated in these two statements:

*<u>Statement 1.</u> For any discrete signal of $N$ samples defined by its $K \le N$ sparse and not necessarily regularly arranged samples, its band-limited, in terms of certain transform $\Phi_N$, approximation defined by Eq. (2.5) can be obtained with mean square error defined by Eq. (2.9) provided positions of the samples secure the existence of the matrix $\mathbf{KofN}_\Phi^{-1}$ inverse to the sub-transform matrix $\mathbf{KofN}_\Phi$ that corresponds to the band-limitation. The approximation error can be minimized by using a transform with the best energy compaction property.*

*<u>Statement 2.</u> Any signal of $N$ samples that is known to have only $K \le N$ non-zero transform coefficients for certain transform $\Phi_N$ ($\Phi_N$ - transform "band-limited" signal) can be fully recovered from exactly $K$ of its samples provided positions of the samples secure the existence of the matrix $\mathbf{KofN}_\Phi^{-1}$ inverse to the sub-transform matrix $\mathbf{KofN}_\Phi$ that corresponds to the band-limitation*

These theorems can be straightforwardly extended to multi-dimensional signals, if band-limitation conditions are separable over signal spectral coordinates as, for instance, in the case when they specify the area of non-zero spectral coefficients in a form of a cube in multi-dimensional space. The case of non-separable band-limitation is more involved and require special analysis in each special case.

## 2.2 Algorithms for signal recovery from sparse non-uniformly sampled data

Implementation of signal recovery from sparse non-uniformly sampled data according to Eq. 2.8 requires matrix inversion, which is, generally, a very computationally demanding procedure. However for some transforms, such as DFT, DCT Walsh, Haar and others that feature fast FFT-type algorithms, pruned versions of these algorithms can be used [2-4]. In many applications one can also be satisfied with signal reconstruction with certain limited accuracy and apply for the reconstruction a simple iterative reconstruction algorithm of the Gershberg-Papoulis [5] type. In this algorithm, the initial guess is generated from available signal samples on a dense sampling grid defined by the accuracy of measuring sample coordinates, supplemented with a guess of the rest of the samples, for which zeros, signal mean value or random numbers can be used. Then, at each iteration, signal is subjected to the selected transform, the transform coefficients are zeroed according to the band-limitation assumption and inverse transformed, after which the next iteration of the restored signal is generated by restoring available signal samples. A review of other iterative and non-iterative algorithmic options one can find in [6].

## 2.3 Analysis of transforms

# Discrete Fourier Transform

Consider the $\mathbf{K}\boldsymbol{of}\mathbf{N}_{DFT}^{LP}$ -trimmed $\boldsymbol{DFT_N}$ matrix:

$$\mathbf{K}\boldsymbol{of}\mathbf{N}_{DFT}^{LP} = \left\{ \mathbf{exp}\left( i2\pi \frac{\tilde{k}\tilde{r}_{LP}}{N} \right) \right\} \tag{2.10}$$

that corresponds to DFT $\mathbf{K}\boldsymbol{of}\mathbf{N}$ -low-pass band-limited signal. Due to complex conjugate symmetry of DFT or real signals, $K$ has to be an odd number, and the set of frequency domain indices of $\mathbf{K}\boldsymbol{of}\mathbf{N}_{DFT}$ low-pass band-limited signals in Eq. (2.10) is defined as:

$$\tilde{r}_{LP} \in \tilde{R}_{LP} = \{[0,1,\ldots,(K-1)/2, N-(K-1)/2,\ldots,N-1]\} \tag{2.11}$$

For such a case, the following theorems hold:

Theorem 1.

$\mathbf{K}\boldsymbol{of}\mathbf{N}$ *-low-pass DFT band-limited signals of **N** samples with only **K** nonzero low frequency DFT coefficients can be precisely recovered from exactly **K** of their samples taken in arbitrary positions.*

Proof.

As it follows from Eqs. (2.3)-(2.8) the theorem is proven if matrix $\mathbf{K}\boldsymbol{of}\mathbf{N}_{DFT}^{LP}$ is invertible. A matrix is invertible if its determinant is nonzero. In order to check whether determinant of the matrix $\mathbf{K}\boldsymbol{of}\mathbf{N}_{DFT}$ is non-zero, permute the order of columns of the matrix as following:

$$\tilde{\tilde{r}} \in \tilde{\tilde{R}} = \{[N-(K-1)/2,\ldots,N-1,0,1,\ldots,(K-1)/2]\} \tag{2.12}$$

and obtain matrix

$$\mathbf{K}\boldsymbol{of}\mathbf{N}_{DFT}^{DFTsh} = \left\{ \mathbf{exp}\left[ i2\pi \frac{\tilde{k}\tilde{\tilde{r}}}{N} \right] \right\} = \left\{ \mathbf{exp}\left[ i2\pi \frac{N-(K-1)/2}{N}\tilde{k} \right] \delta\left( \tilde{k} - \tilde{\tilde{r}} \right) \right\} \left\{ \mathbf{exp}\left[ i2\pi \frac{\tilde{k}\tilde{\tilde{\tilde{r}}}}{N} \right] \right\} \tag{2.13}$$

where

$$\tilde{\tilde{\tilde{r}}} \in \tilde{\tilde{\tilde{R}}} = \{[0,\ldots,K-1]\} \tag{2.14}$$

The first matrix in this product of matrices is a diagonal matrix, which is obviously invertible. The second one is a version of Vandermonde matrices, which are also known to have non-zero determinant if, like in our case, its ratios for each row are distinct ([7])

As permutation of the matrix columns does not change the absolute value of its determinant, Eq. (2.13) implies that determinant of $\mathbf{K}\boldsymbol{of}\mathbf{N}$ -trimmed $\boldsymbol{DFT_N}$ matrix of Eq. (2.10) is also non-zero for arbitrary set $\tilde{K} = \{\tilde{k}\}$ of positions of $K$ available signal samples.

One can easily see that for DFT $\mathbf{K}\boldsymbol{of}\mathbf{N}$ -high-pass band-limited signals, for which

$$\mathbf{KofN}_{DFT}^{HP} = \left\{ \exp\left( i2\pi \frac{\tilde{k}\tilde{r}_{HP}}{N} \right) \right\} \qquad (2.15)$$

where

$$\tilde{r}_{HP} \in \tilde{R}_{HP} = \{[(N-K+1)/2, (N-K+3)/2, .., (N+K-1)/2]\} \qquad (2.16)$$

a similar theorem holds

<u>Theorem 2</u>.

***KofN** -high-pass DFT band-limited signals of **N** samples with only **K** nonzero high frequency DFT coefficients can be precisely recovered from exactly **K** of their arbitrarily taken samples.*

Note that, due to the complex conjugate symmetry of DFT of real signals, $K$ in this case has to be odd whatever $N$ is.

Obviously, above Theorems 1 and 2 can be extended to a more general case of signal DFT band limitation, when indices $\{\tilde{r}\}$ of nonzero DFT spectral coefficients form arithmetic progressions with common difference other than one such as, for instance,

$$\tilde{r}_{mLP} \in \tilde{R}_{mLP} = \left\{ 0, m, ..., m\frac{(K-1)}{2}, N - m\frac{(K-1)}{2}, ..., N - m\frac{(K-1)}{2} + \frac{(K+1)}{2} \right\} \qquad (2.17)$$

An illustrative example of DFT-band-limited signal recovery from sparse samples taken in random positions and within a compact interval is presented in Fig. 1.

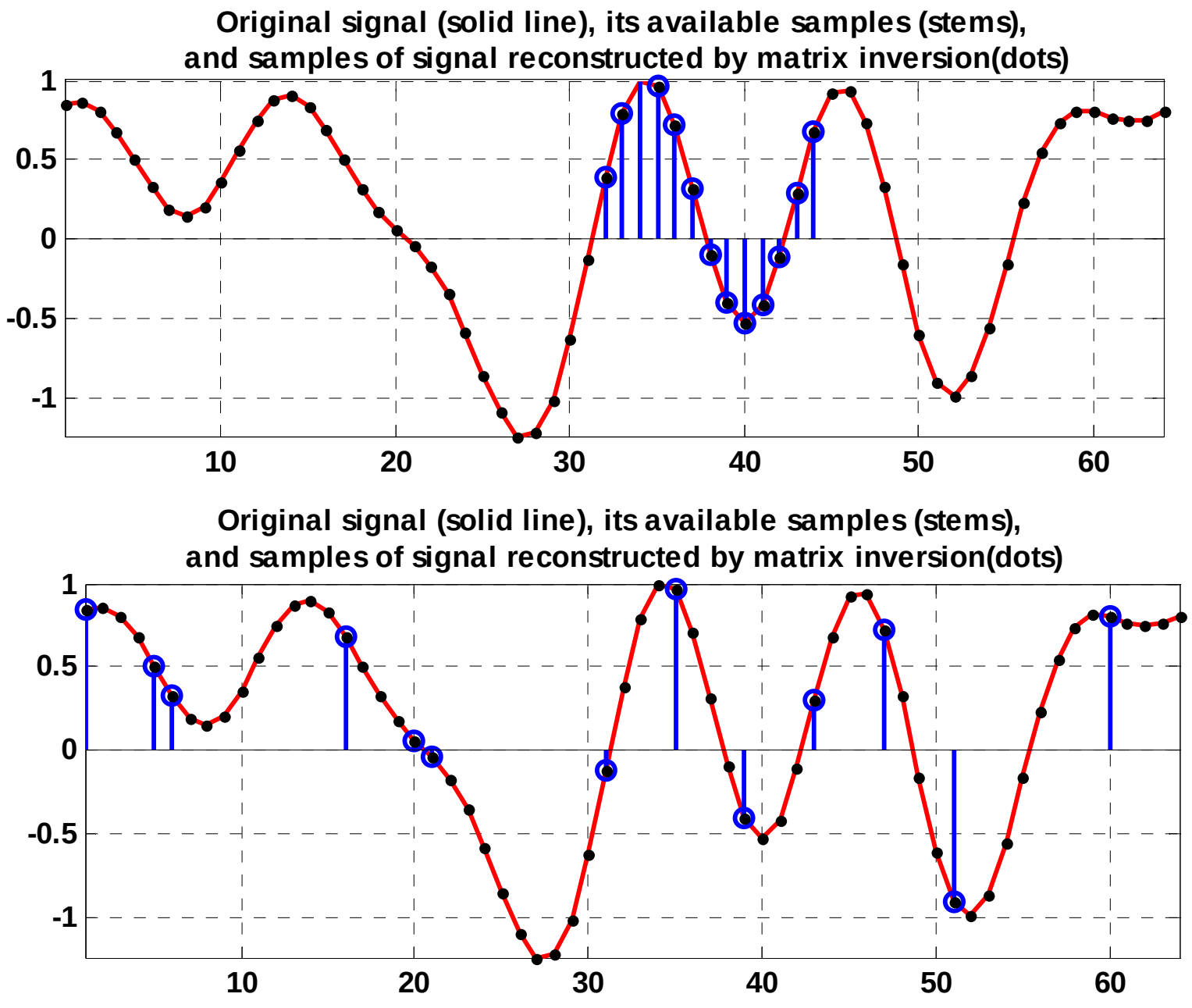

Fig. 1. Recovery, by means of matrix inversion, of a DFT-band-limited pseudo-random signal from sparse samples taken in random positions (upper plot) and within a compact interval (bottom plot

## Discrete Cosine Transform (DCT)

$N$-point Discrete Cosine Transform of a signal is equivalent to $2N$-point Shifted Discrete Fourier Transform (SDFT) with shift parameters (1/2,0) of $2N$- sample signal obtained from the initial one by its mirror reflection [4]. $\mathbf{K}of\mathbf{N}$ -trimmed matrix of SDFT(1/2,0)

$$\mathbf{K}of\mathbf{N}_{SDFT} = \left\{ \exp\left( i2\pi \frac{(\tilde{k}+1/2)\tilde{r}}{2N} \right) \right\} \tag{2.18}$$

can be represented as a product

$$\mathbf{K}of\mathbf{N}_{SDFT} = \left\{ \exp\left( i2\pi \frac{\tilde{k}\tilde{r}}{2N} \right) \left\{ \exp\left( i\pi \frac{\tilde{r}}{2N} \right) \delta(k-r) \right\} \right\} = \mathbf{K}of\mathbf{N}_{DFT} \left\{ \exp\left( i\pi \frac{\tilde{r}}{2N} \right) \delta(k-r) \right\} \tag{2.19}$$

of a $2N$ -point DFT matrix and a diagonal matrix $\{\exp(i\pi\tilde{r}/2N)\delta(k-r)\}$. The latter one is invertible and the invertibility of $\mathbf{K}of\mathbf{N}$ -trimmed $DFT_{2N}$ matrix $\mathbf{K}of\mathbf{N}_{DFT}$ can be proved, for above described band-limitations, as it was done above for the DFT case. Therefore, for DCT theorems similar to those for DFT hold.

These theorems hold also for 2D DFT and DCT transforms provided band-limitation conditions are separable. The case of non-separable band-limitation requires further study. Note that working in DFT or DCT domain results, in the case of low-pass band-limitation, in signal discrete sinc-interpolation [4].

## Discrete Fresnel transform

Canonical Discrete Fresnel Transform is defined as ([4])

$$a_k = \frac{1}{\sqrt{N}} \sum_{r=0}^{N-1} \alpha_r \exp\left[ -i\pi \frac{(k\mu - r/\mu)^2}{N} \right] \tag{2.20}$$

It can easily be expressed via DFT:

$$\alpha_r = \frac{1}{\sqrt{N}} \left\{ \sum_{k=0}^{N-1} \left[ a_k \exp\left( i\pi \frac{k^2\mu^2}{N} \right) \right] \exp\left( -i2\pi \frac{kr}{N} \right) \right\} \exp\left( i\pi \frac{r^2}{\mu^2 N} \right) \tag{2.21}$$

In a matrix form, it can be represented as a matrix product of diagonal matrices and the matrix of Discrete Fourier Transform. Therefore for Discrete Fresnel Transform formulation of band-limitation and requirements to positions of sparse samples are the similar to those for DFT.

## Wavelets and Other Bases

The main peculiarity of wavelet bases is that their basis functions are most naturally ordered in terms of two components: scale and position within the scale. Scale index is analogous to the frequency index for DFT. Position index tells only of the shift of the same basis function within the signal on each scale. Therefore, band-limitation for DFT translates to scale limitation for wavelets. Limitation in terms of position is trivial: it simply means that some parts of the signal are not relevant. Commonly, discrete wavelets are designed for signals whose length is an integer power of 2 ( $N = 2^n$ ). For such signals, there are $s \le n$ scales and possible "band-limitations".

The simplest special case of wavelet bases is Haar basis. Signals with $N = 2^n$ samples and only with $K$ lower index non-zero Haar transform (the transform coefficients $\{K,...,N-1\}$ are zero) are ($\tilde{s} = \left(\lfloor \log_2 (K-1) \rfloor + 1\right)$) - “band-limited”, where $\lfloor x \rfloor$ is an integer part of $x$. Such signals are piecewise constant within intervals between basis function zero-crossings. The shortest intervals of the signal constancy contain $2^{n-\tilde{s}}$ samples. As one can see from Fig. 2 (a), for any two samples that are located on the same interval, all Haar basis function on this and lower scales have the same value. Therefore, having more than one sample per constant interval will not change the rank of the matrix **KofN**. The condition for perfect reconstruction is, therefore, to have at least one sample on each of those intervals.

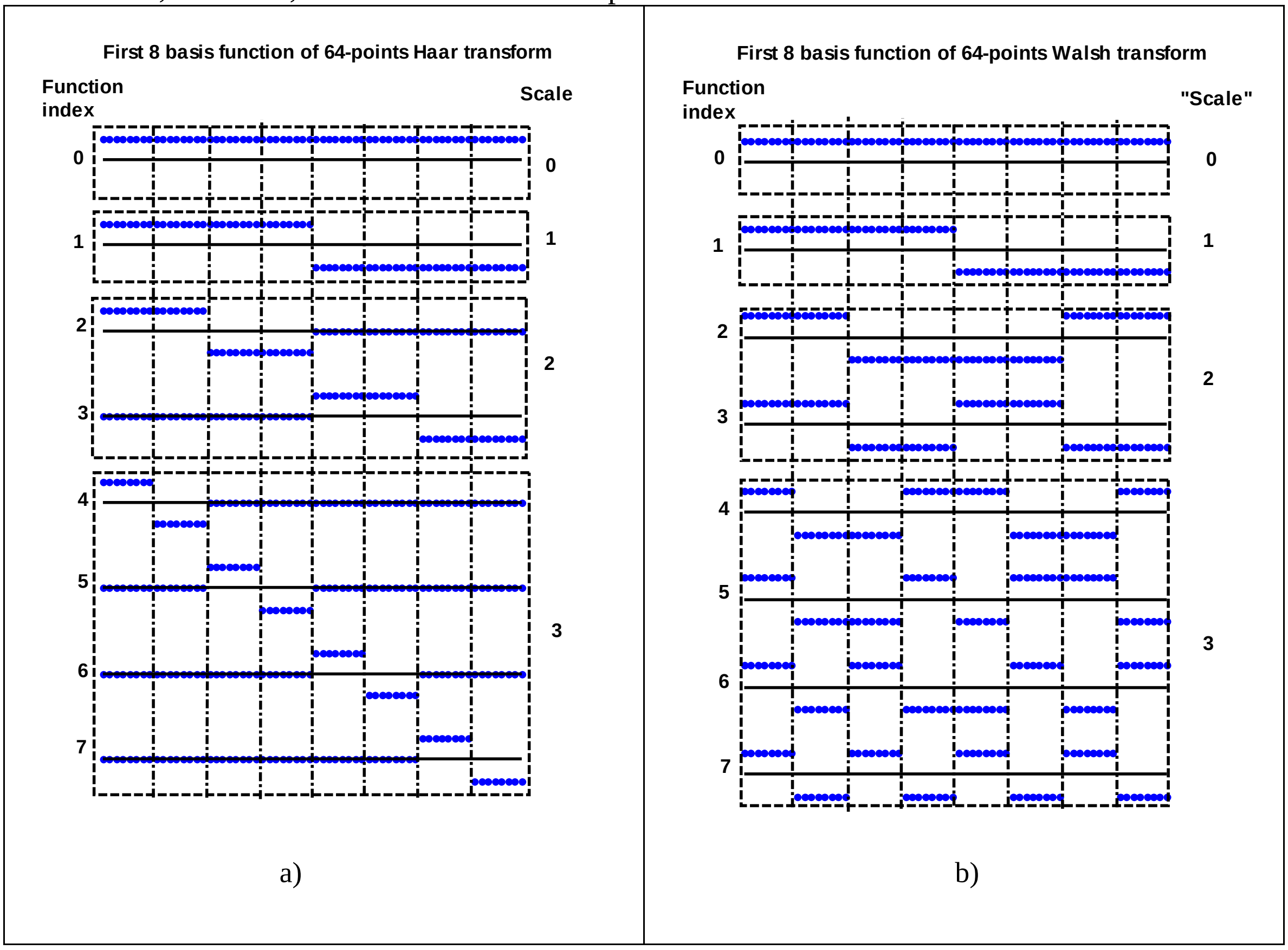

Fig. 2. Haar (a) and Walsh (b) basis functions

For other wavelets as well as for other bases a general necessary, sufficient and easily verified condition for the invertibility of **KofN**-trimmed transform sub-matrix is not known for the present authors. Standard linear algebra procedures for determining matrix rank can be used for testing invertibility of the matrix.

For Walsh basis functions, the index corresponds to the “sequency”, or to the number of zero crossings of the basis function. The sequency carries a certain analogy to the signal frequency. Basis functions ordering according to their sequency, which is characteristic for Walsh transform, preserves, for many real-life signals, the property of decaying transform coefficients’ energy with their index. Therefore, for Walsh transform the notion of low-pass band-limited signal approximation, similar to the one described for DFT, can be used. On the other hand, as one can see from Fig. 2 (b), Walsh basis functions, similarly to Haar basis function, can be characterized by the scale index, which specifies the shortest interval of signal constancy. Signals with $N = 2^n$ samples and band-limitation of $K$ Walsh transform coefficients have shortest intervals of signal constancy of $2^{n-\tilde{s}}$ samples, where $\tilde{s} = \left(\lfloor \log_2 (K-1) \rfloor + 1\right)$. A necessary condition for perfect reconstruction is to have $K$ signal samples taken on different intervals. Unlike the Haar transform case, not all the intervals are needed to be sampled, but only $K$ intervals out off the total number of intervals. For a special case of $K$ equal to a power of 2, there are $K$ intervals, each of which has to be sampled to secure perfect reconstruction, This is the case, when the reconstruction condition for Walsh Transform is identical to that for Haar transform.

Fig. 3 illustrates the case of recovery of an image “band limited” in the Haar transform domain. Two examples are shown: arrangement of sparse samples, for which signal recovery is possible (a) and that for which signal is not recoverable (b). Note that when the Haar reconstruction is possible, it is reduced to the trivial nearest neighbor interpolation.

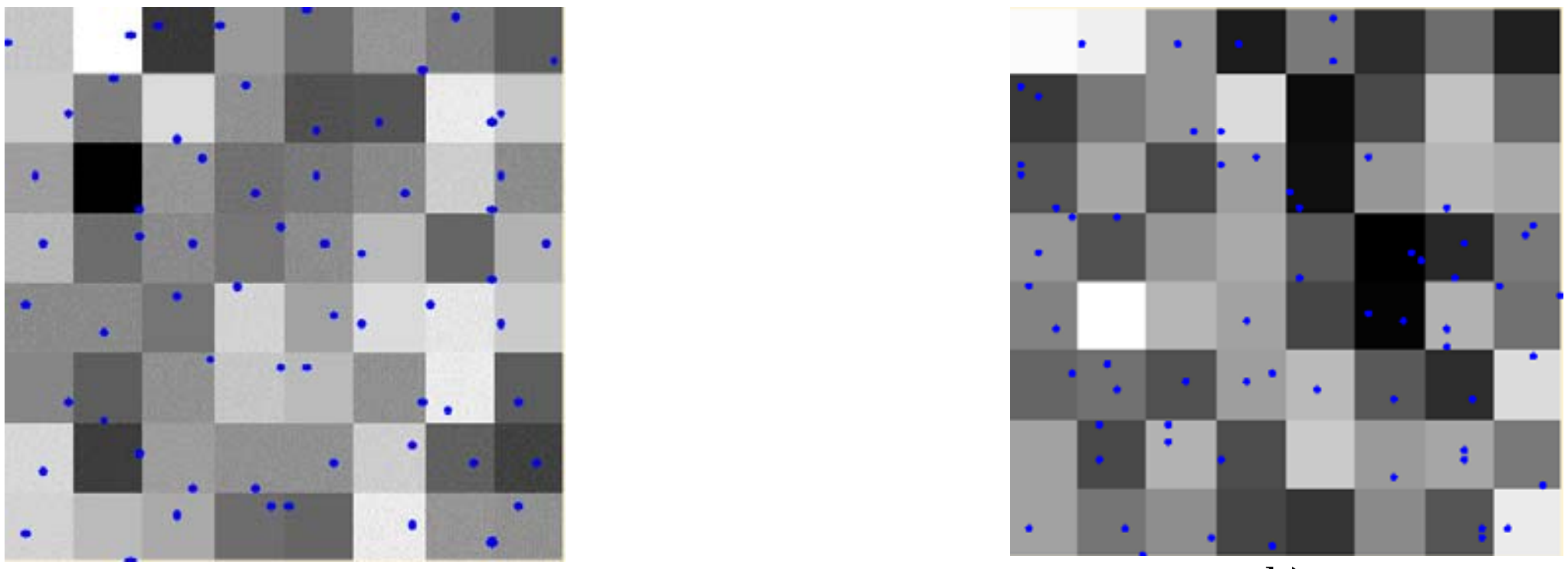

a) b)

Fig. 3. Recoverable (a) and non-recoverable (b) arrangements of image sparse samples (shown by dots)

An example of perfect reconstruction of Walsh transform domain “band-limited” signal of $\boldsymbol{N}$=512 and band limitation $\boldsymbol{K}$=5 is illustrated in Fig. 4. In this example, the resulted $\boldsymbol{KofN}^{\boldsymbol{Walsh}}$ matrix is:

$$\boldsymbol{KofN}^{\boldsymbol{Walsh}}\Big|_{K=5} = \begin{bmatrix} \mathbf{1} & \mathbf{-1} & \mathbf{1} & \mathbf{-1} & \mathbf{-1} \\ \mathbf{1} & \mathbf{-1} & \mathbf{-1} & \mathbf{1} & \mathbf{1} \\ \mathbf{1} & \mathbf{1} & \mathbf{1} & \mathbf{1} & \mathbf{1} \\ \mathbf{1} & \mathbf{1} & \mathbf{-1} & \mathbf{-1} & \mathbf{-1} \\ \mathbf{1} & \mathbf{1} & \mathbf{1} & \mathbf{1} & \mathbf{-1} \end{bmatrix}$$

and its rank equals to 5. One should note that, in this particular example, perfect reconstruction in the Haar transform domain is not possible since one of the shortest intervals of the signal constancy contains no samples.

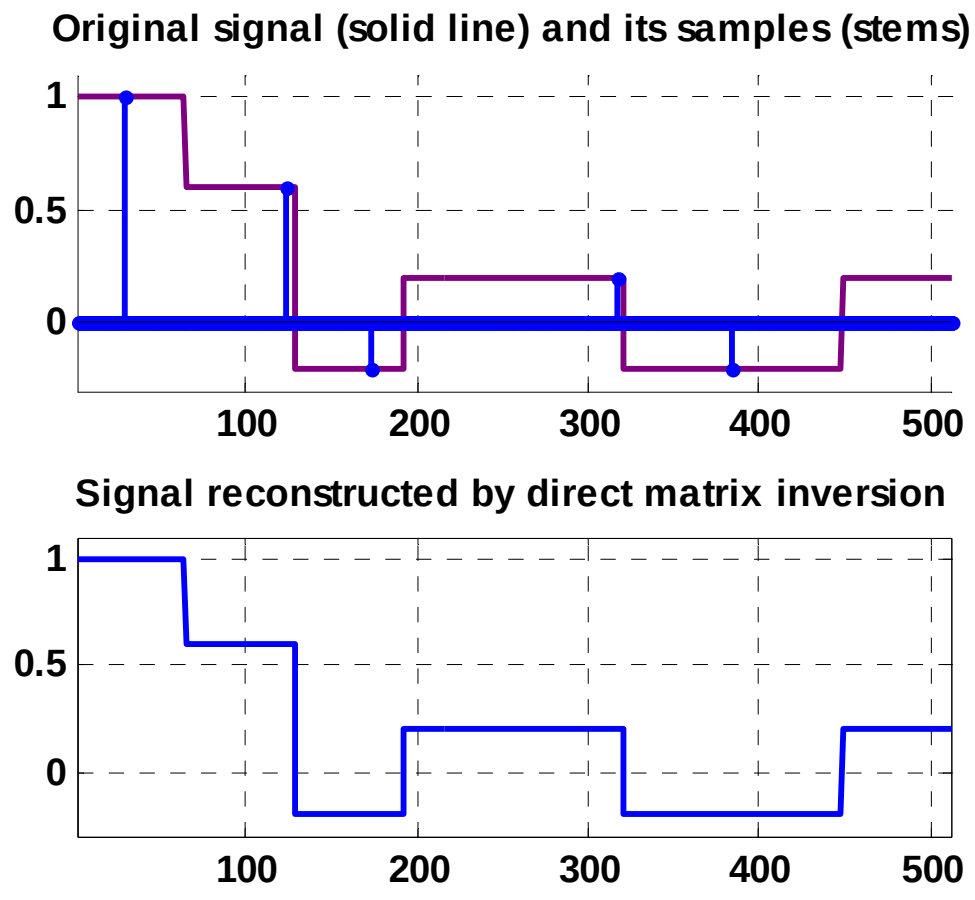

Fig. 4. An example of Walsh-band-limited signal recovery by means of matrix inversion

## 2.4 Signal recovery from sparse or non uniformly sampled data as an approximation task

There might be numerous applications of above described algorithms for recovery band-limited signals and generating band-limited approximations of signals from sparse data. For the latter, the above theory and algorithms can be applied as following:

1. Given a certain number of available signal samples, specify, according to the accuracy with which physical coordinates are known or using other a priori data, the signal dense sampling grid and the required number of samples to be recovered.
2. Select a transform with presumably better energy compaction capability for the signal at hand and specify the signal band limitation in the domain of this transform.
3. Place available signal samples on the signal dense sampling grid and run the direct matrix inversion or iterative reconstruction algorithm.

We illustrate possible applications on two examples.

One of attractive potential applications of the above signal recovery technique is image super-resolution from multiple video frames with chaotic pixel displacements due to atmospheric turbulence, camera instability or similar random factors ([9]). By means of elastic registration of sequence of frames of the same scene, one can determine, for each image frame and with sub-pixel accuracy, pixel displacements caused by random acquisition factors. Using these data, a synthetic fused image can be generated by placing pixels from all available video frames in their proper positions on the correspondingly denser sampling grid according to their found displacements. In this process, some pixel positions on the denser sampling grid will remain unoccupied, especially when limited number of image frames are fused. These missing pixels can then be restored using the above described iterative band-limited interpolation algorithm. Computer simulation reported in ([9]) showed that application of the iterative interpolation may substantially improve results of image resolution enhancement by fusing multiple frames with different local displacements. This is illustrated in Fig. 5, which shows one low resolution frame (a), image fused from 30 frames (b) and a result of iterative interpolation (c) achieved after 50 iterations. Image band limitation was set in this experiment twice of the base band of raw low resolution images.

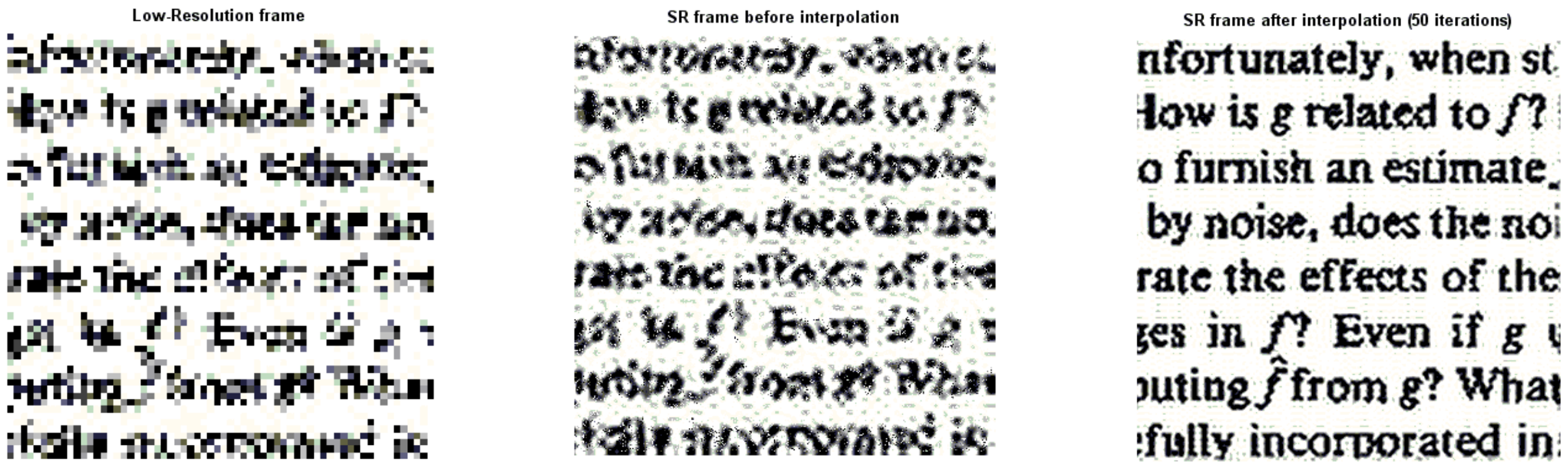

a) b) c)

Fig. 5. Iterative image interpolation in the super-resolution process: a) – a low resolution frame; b) image fused by elastic image registration from 50 frames; c) – a result of iterative interpolation of image b) after 50 iterations.

Another straightforward application the discussed sparse data recovery algorithm can find in tomographic imaging, where it frequently happens that a substantial part of slices, which surrounds the body slice, is known to be an empty field. This means that slice projections (sinograms) are Radon transform "band-limited" functions. Therefore whatever number of projections is available, a certain number of additional projections, commensurable, according to the discrete sampling theorem, with the size of the slice empty zone, can be obtained and the corresponding resolution increase in the reconstructed images can be achieved using the described iterative band-limited reconstruction algorithm. Fig. 6 illustrates such super-resolution by means of recovery of missing half of projections achieved using the fact that by simple segmentation of the restored image shown in Fig.5 (b) it was found that the outer 55%  of the image area is empty.

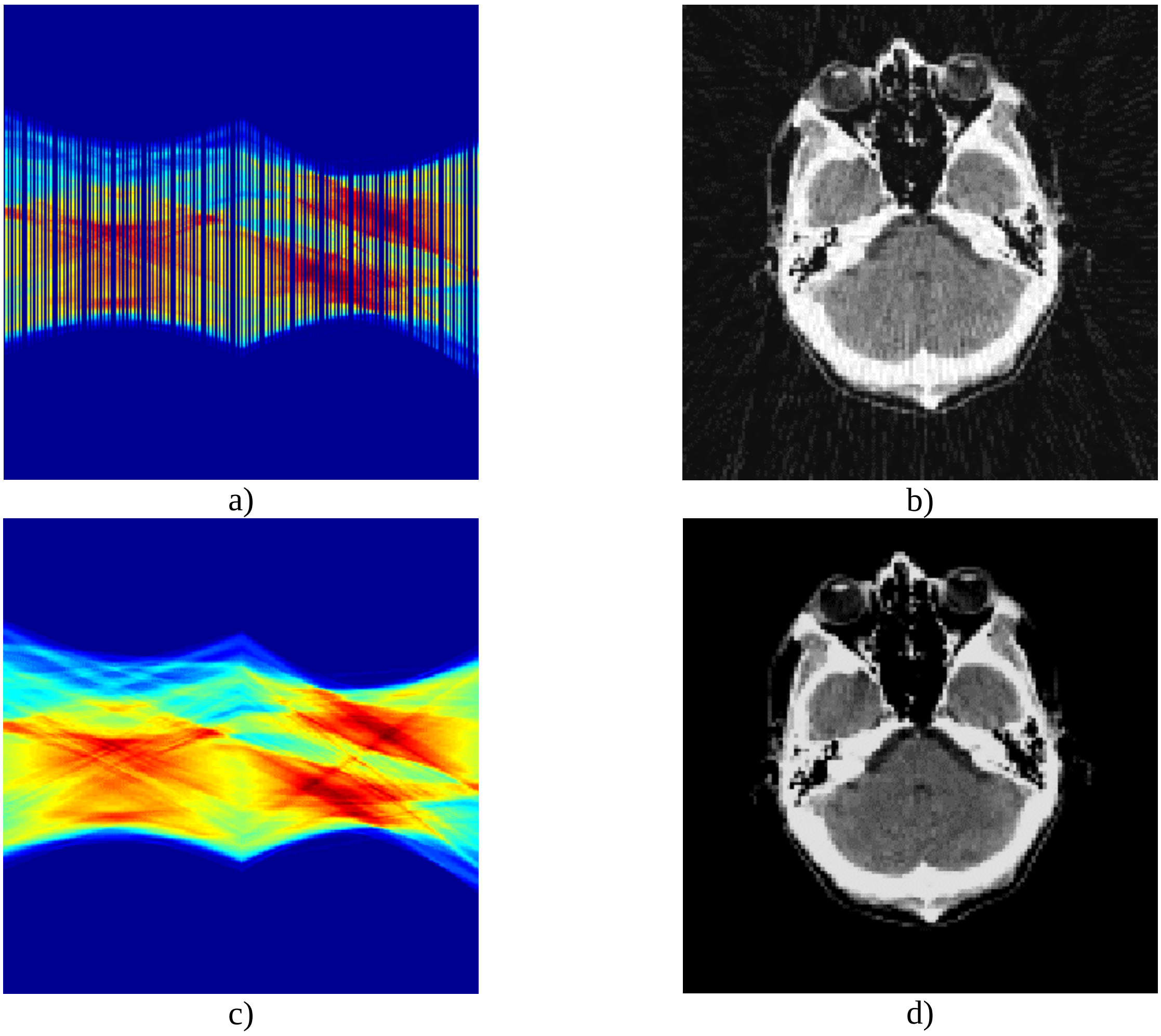

Fig. 6. Super-resolution in computed tomography: a) – set of initial projections supplemented with the same number of presumably lost projections to double the number of projections; initial guesses of the supplemented projections are set to zero; b) image reconstructed from initially available projections; c) result of iterative restoration of missing projections; d) – image reconstructed from the restored double set of projections

## 3. Sharply Band Limited Discrete Signals with Sharply Limited Support and the Discrete Uncertainty Principle

It is well known that continuous signals can't be both strictly band-limited and have strictly bounded support. In fact, continuous signals are neither band-limited nor have strictly bounded support. They can only be more or less densely concentrated in signal and spectral domains. This property is mathematically formulated in the form of the "uncertainty principle":

$$X_{\varepsilon S} \times F_{\varepsilon B} > 1 \ , \qquad (3.1)$$

where $X_{\varepsilon S}$ is interval in signal domain that contains $(1-\varepsilon S)$- fraction of its entire energy, $F_{\varepsilon B}$ is interval in signal Fourier spectral domain that contains $(1-\varepsilon B)$- fraction of signal energy and both $\varepsilon S$ and $\varepsilon B$ are assumed to be sufficiently small.

In distinction to that, sampled signals that represent continuous signals can be sharply bounded both in signal and spectral domains specified by the finite number of signal samples. This is quite obvious for some signal spectral presentation such as Haar signal spectra. In particular, Haar basis functions are examples of sampled functions sharply bounded in signal and Haar spectral domains. But it turns out that this property of signal sharp boundedness in both signal and spectral domains holds also for Discrete Fourier Transform and Discrete Cosine Transform spectra, which are discrete representation of Fourier integral transform. Such space-frequency sharply bounded signals can be generated using the above-described iterative algorithm that, at each iteration, applies requested bounds alternatively in signal and spectral domains. Examples of such space-frequency sharply bounded images are shown in Fig. 3.1.

Relationship between bounds in signal and DFT domains is defined by the discrete uncertainty principle. The discrete uncertainty principle can be derived from the continuous one using the same reasoning that are used in signal sampling. Let $N_{sign}$ be the number of signal non-zero samples, $N_{spectr}$ be the number of its non-zero spectral samples and $N$ be the number of samples in the signal sampling grid. Then the length of the continuous signal that corresponds to the given sampled signal can be estimated as $X \approx N_{sign}\Delta x$, where $\Delta x$ is the signal sampling interval and length of the interval in signal Fourier domain occupied by the signal spectrum can be estimated as $F \approx N_{spectr}\Delta f$, where $\Delta f$ is signal spectrum sampling interval. From the uncertainty principle for continuous signals (Eq. 3.1) it follows that

$$N_{sign}\Delta x N_{spectr}\Delta f > 1, \tag{3.2}$$

or

$$N_{sign}N_{spectr} > \frac{1}{\Delta x \Delta f} \tag{3.3}$$

As, according to the cardinal sampling relationship, for which DFT represents integral Fourier Transform,

$$\frac{1}{\Delta x \Delta f} = N, \tag{3.4}$$

we obtain finally that

$$N_{sign}N_{spectr} > N, \tag{3.5, a}$$

or

$$N_{sign}N_{spectr} > cN \tag{3.5, b}$$

where $c > 1$ is a certain constant. The relationships (3.5) formulate the discrete uncertainty principle. The discrete uncertainty principle was, apparently for the first time, addressed in Ref.10.

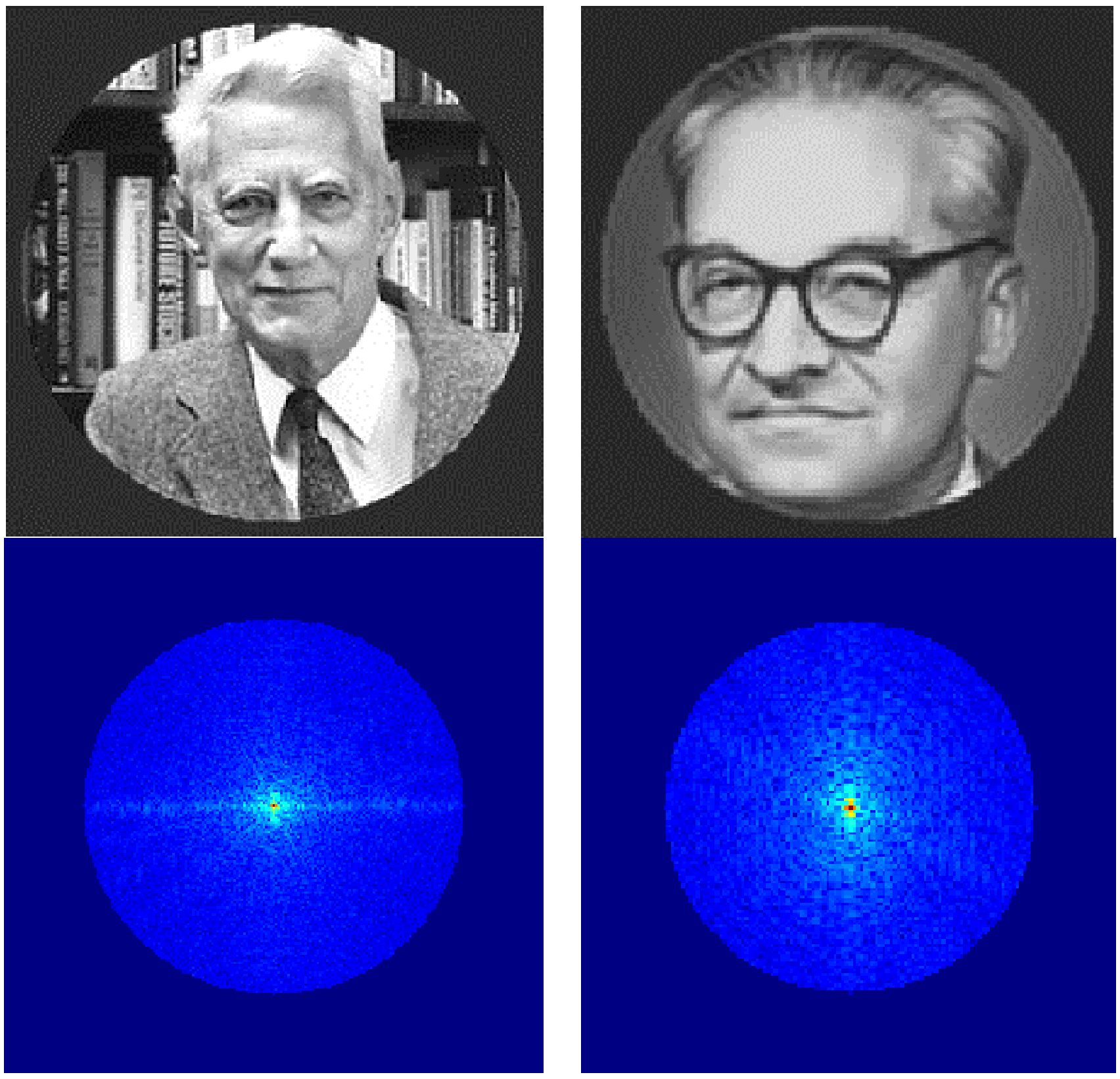

Fig. 7. Pictures of C. Shannon (left, upper row) and V. Kotelnikov (right, upper row) and their corresponding DFT spectra bounded by circular windows (DFT spectra are shown centered at their DC-component)

## 4. Sinc-lets: Sharply Band Limited Basis Functions with Sharply Limited Support

The existence of sharply space-frequency bounded signals allows to suggest the existence of a family of correspondingly sharply space-frequency bounded basis functions that can be used to represent such signals. Although analytical representation of these functions is yet to be found, one can generate them using the same iterative procedure for generation sharply space-frequency bounded signals using as seed signals delta-functions with different locations within the selected interval of the signals in signal domain. We call these functions sinclets to reflect the fact that they resemble windowed discrete sinc-functions as one can see from illustrative examples presented in Fig. 8. For generating these function the following algorithm was used:

$$\left[\mathbf{sinclet}_{S\lim}^{B\lim}(k)\right]^{(0)} = \delta(k-k_0),\ \mathbf{sinclet}_{S\lim}^{B\lim}(k) = \lim_{t\to\infty}\left[\mathbf{sinclet}_{S\lim}^{B\lim}(k)\right]^{(t)}, \qquad (4.1)$$

where

$$\left[\mathbf{sinclet}_{S\lim}^{B\lim}(k)\right]^{(t)} = S_{\lim} \bullet IDFT\left\{B_{\lim} \bullet DFT\left\{\left[\mathbf{sinclet}_{S\lim}^{B\lim}(k)\right]^{(t-1)}\right\}\right\}, \qquad (4.2)$$

$S_{\lim}$ and $B_{\lim}$ are operators of signal band limitation and space limitations, correspondingly, $k_0$ -is index of position of the function. This algorithm can be regarded as an algorithmic definition of sinc-lets. In what follows we will consider only low-pass band-limited sinclets (LP-sinclets).

As one can see from Fig. 8, LP sinc-lets are shift variant functions: their shape and height depend on the position.

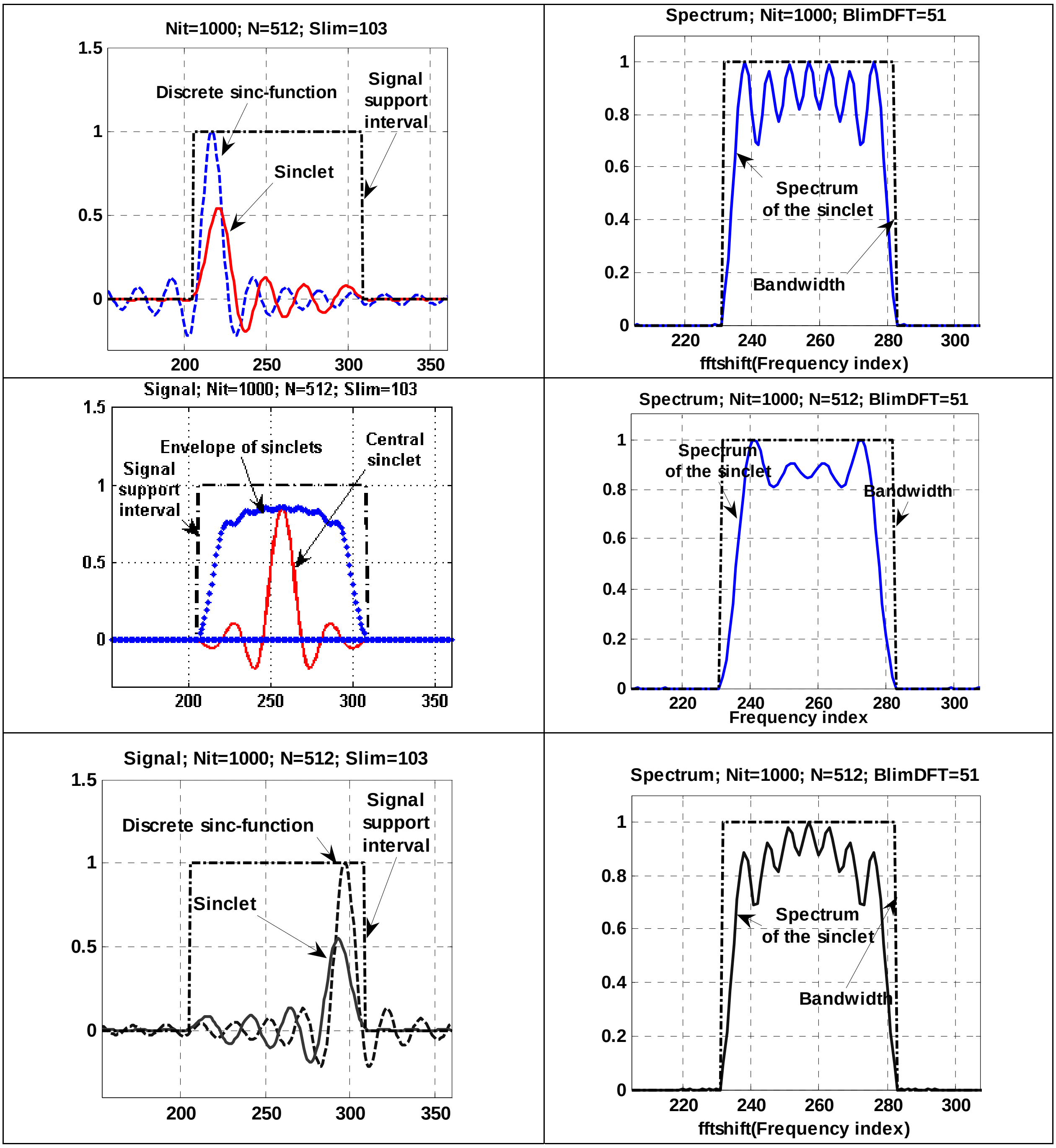

Fig. 8. Examples of LP sinc-lets in different positions within interval of 103 samples for a signal of 512 samples (left column) and their corresponding DFT spectra (right column). LP sinc-lets are shown along with the corresponding discrete sinc-functions of the same bandwidth

Fig. 9. illustrates the speed of convergence of generated signals to their fixed point

$$\left[\mathrm{sinclet}_{S\lim}^{B\lim}(k)\right] = \left\{ S_{\lim} \bullet IDFT\left\{ B_{\lim} \bullet DFT\left\{ \left[\mathrm{sinclet}_{S\lim}^{B\lim}(k)\right] \right\}\right\}\right\} \qquad (4.3)$$

for cases of space interval of 103 samples and spectral interval of 51 and 103 samples out of 512. On vertical axes on these plots, fraction of the signal energy outside the selected bounded interval, in this particular case interval of 103 samples out of 512.

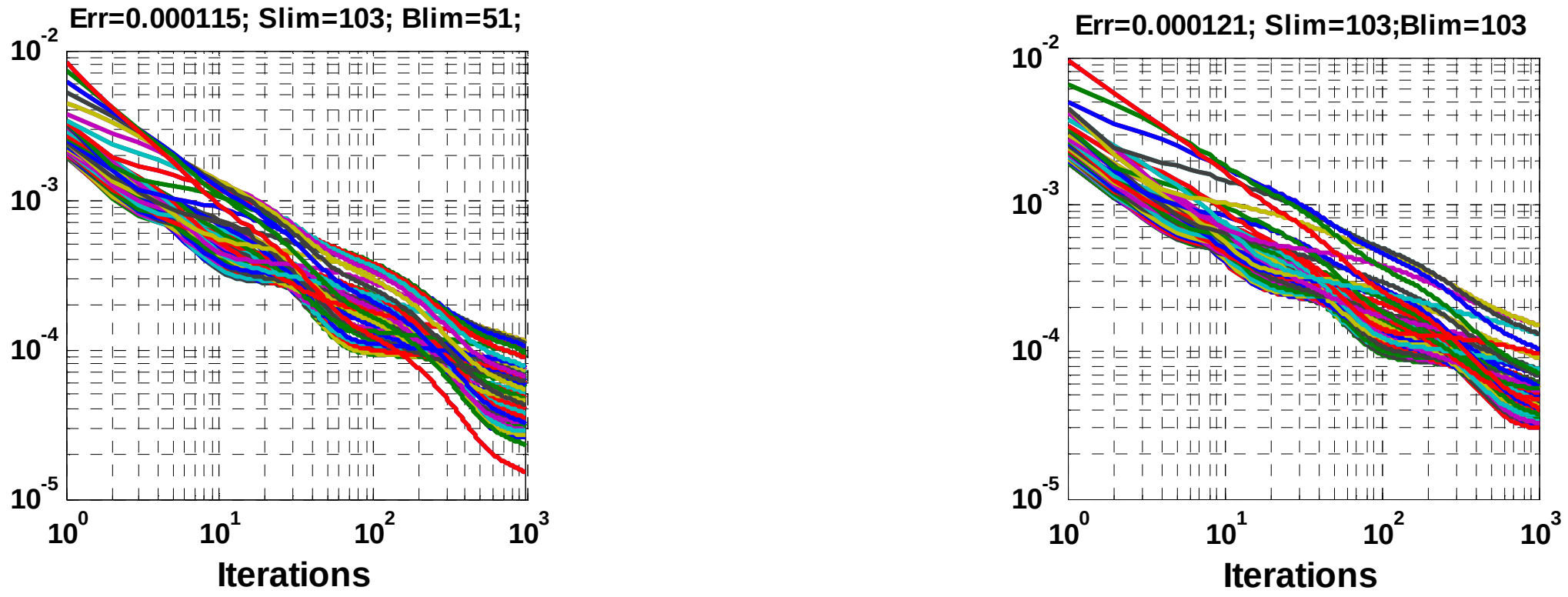

Fig. 9. Illustrative example of iteration convergence in generation LP sinc-lets shown in Fig. 8 (left) and sinc-lets obtained on sampling grid of 512 samples for 103 non-zero signal samples and 103 non-zero samples of their DFT spectrum. Each particular curve on these plots corresponds to different positions of sinc-lets. Vertical axes on the plots represent fraction of the signal energy outside the selected interval of 103 samples.

Fig. 10 shows matrices of mutual correlations of LP sinc-lets in different positions obtained for space interval of 103 samples and spectral interval of 51 and 103 samples. These matrices allow to hypothesize that sinc-lets shifted by interval $\Delta N = N/B\lim$ inversely proportional their bandwidth interval $B\lim$ form a family of orthogonal functions.

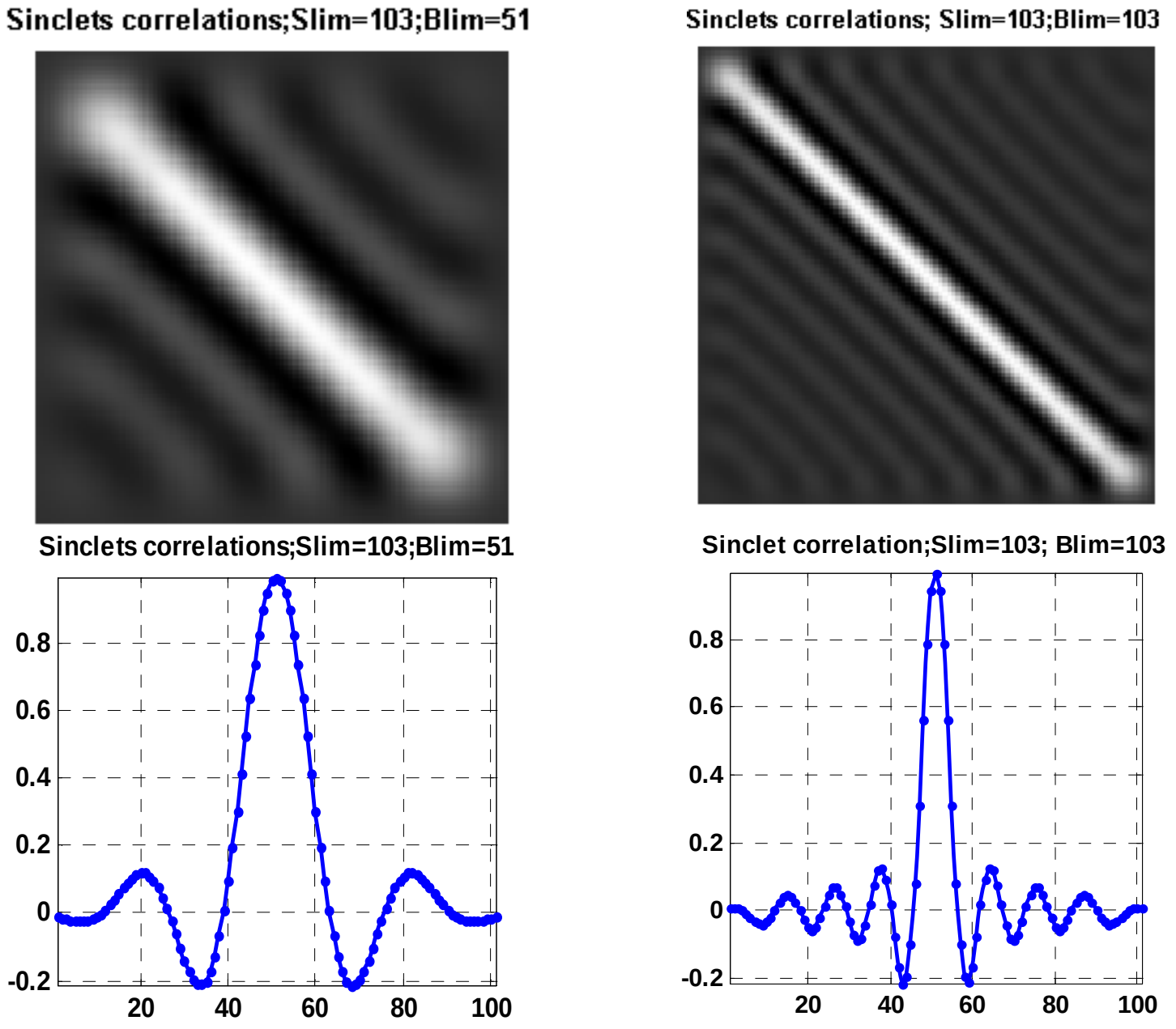

Fig. 10. Matrices of mutual correlations of LP sinc-lets in different positions for two families of sinc-lets obtained for space limitation of 103 samples and band-limitation of 51 and 103 samples (upper row) and their corresponding central sections (bottom row)

Band limitation in DCT domain generates similar sinc-lets, as it is illustrated in Fig. 11. The shape of two dimensional LP sinclets depends on the shape of their space and spectrum limitation. Obviously, for separable space and spectrum limitation 2D sinclets are products of corresponding 1-D sinclets. Examples of 2D sinclets limited in space by circles and square and circularly limited in DFT and DCT domains are shown in Fig. 12.

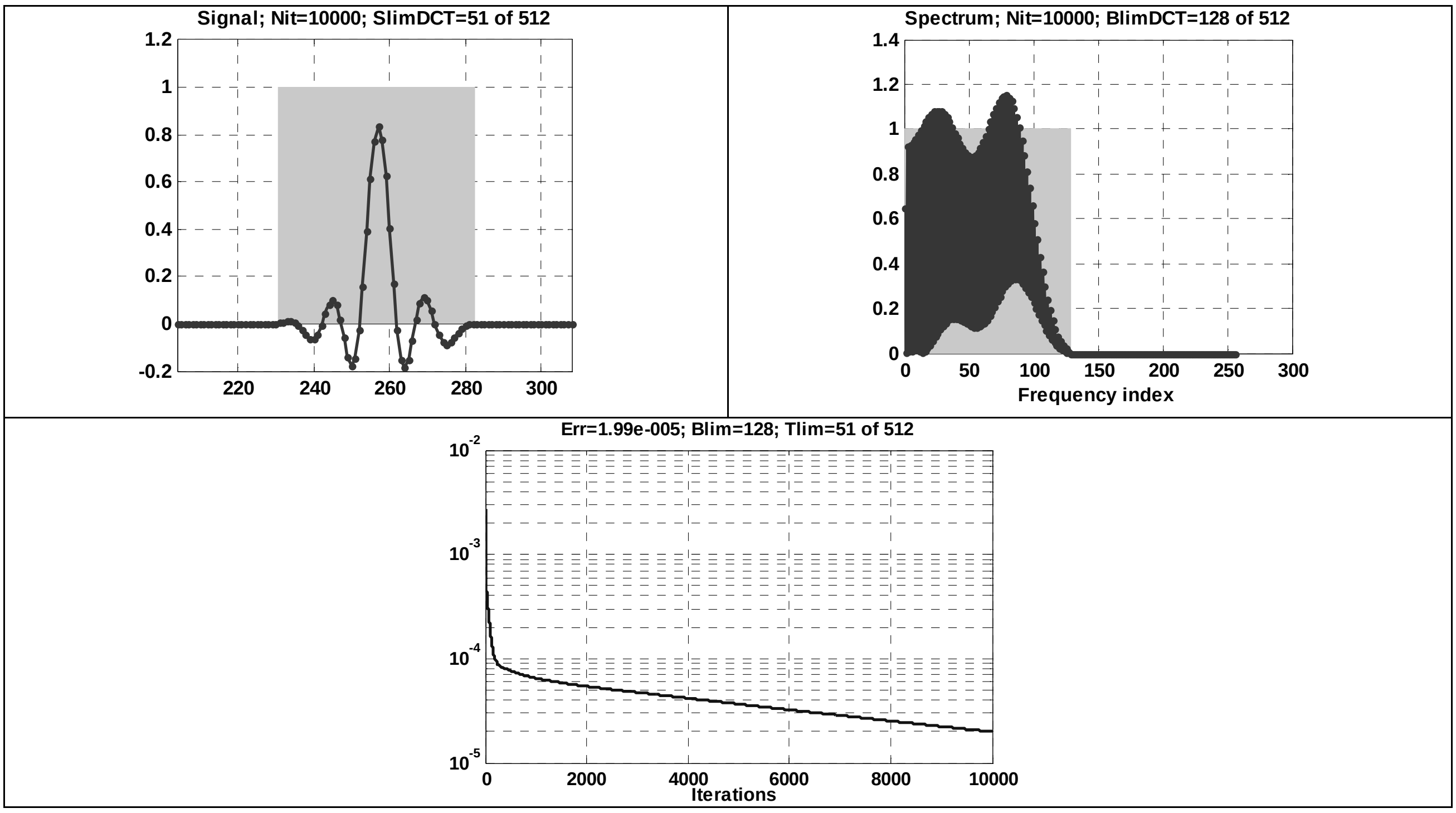

Fig. 11. Sinc-let (upper left) and its DCT spectrum (upper right) generated by band-limitation in DCT domain and plot of signal residual energy outside the selected interval versus the number of iterations (bottom). Gray rectangles indicate selected intervals in space and frequency domains

## 5. Conclusion

In the paper, we formulated the Discrete sampling theorem that refers to discrete signals that are band-limited in a domain of a certain transform and states that “***KofN*** band-limited” discrete signals of ***N*** samples can be precisely recovered from their ***K*** sparse samples provided positions of the available samples satisfy certain limitations, which depend on the transform. This theorem provides a tool for optimal, in terms of root mean squared error, approximation of arbitrary discrete signals specified by their sparse samples with “***KofN***- band-limited” signals, provided appropriate selection of the signal representation transform. Properties of different transforms, such as Discrete Fourier, Discrete Cosine, Discrete Fresnel Transform, Haar, Walsh and wavelet transforms, relevant to application of the Discrete Sampling Theorem are discussed and, in particular, it is shown that precise reconstruction of one-dimensional “***KofN***-DFT band-limited” and “***KofN***-DCT band-limited” signals is always possible from sparse samples regardless of sample positions on the signal dense grid and that same holds for two-dimensional signals provided separable band-limitation conditions. The case of non-separable band limitation turned to be of more involved nature and requires further study.

Applications of the discrete sampling theorem based approach to image recovery from sparse data are illustrated on examples of image super-resolution from multiple chaotically sampled frames and image reconstruction from sparsely sampled projections. For the latter case, it is shown that, in applications, where object slices contain areas, which a priori are known to be empty, reconstruction of slice images from a given set of projections is possible with super-resolution.

We also provided an experimental evidence of the existence of sharply space- band-limited in DFT and DCT domain signals provided the number of their non-zero samples $N_{sign}$ and their non-zero spectral samples $N_{spectr}$ satisfy the discrete uncertainty principle given by Eq. 3.5. Finally, we introduced a new family of sharply space and band-limited basis function “sinc-lets” and illustrated their appearance and properties using experimental data.

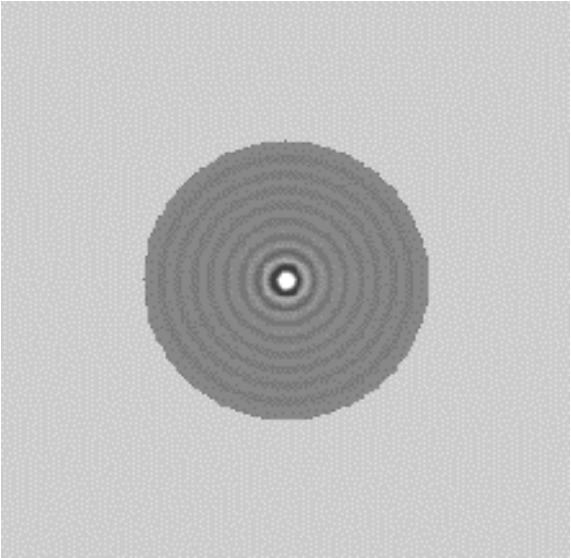

LP DFT-Sinclets: N=256x256;SzLim=0.5; Blim=0.15

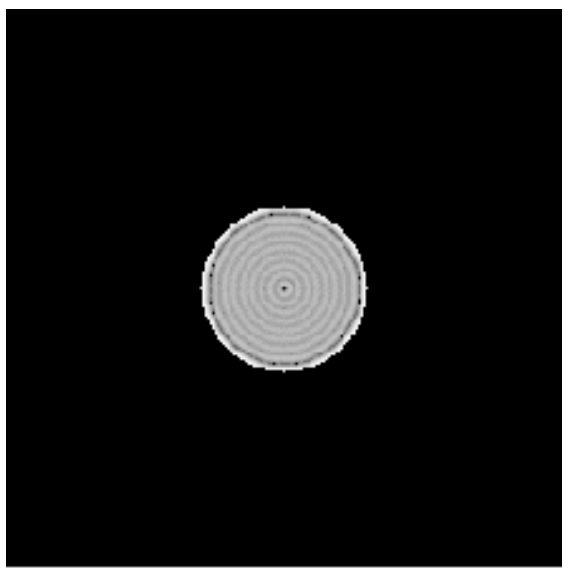

LP DFT-Sinclet spectrum:N=256x256;SzLim=0.5; Blim=0.15

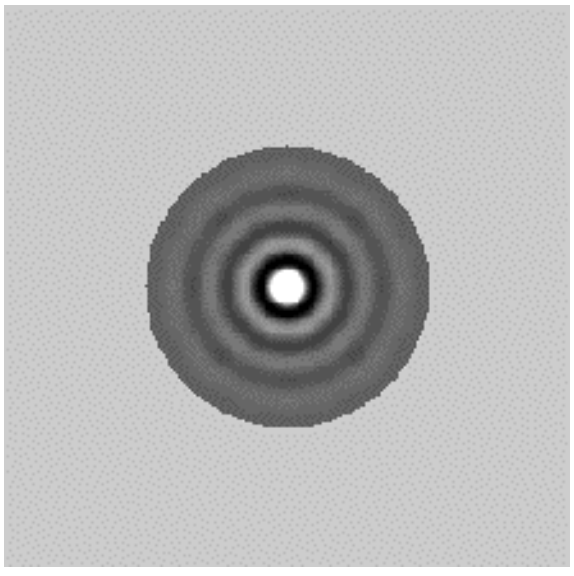

LP DFT-Sinclets: N=256x256;SzLim=0.5; Blim=0.15

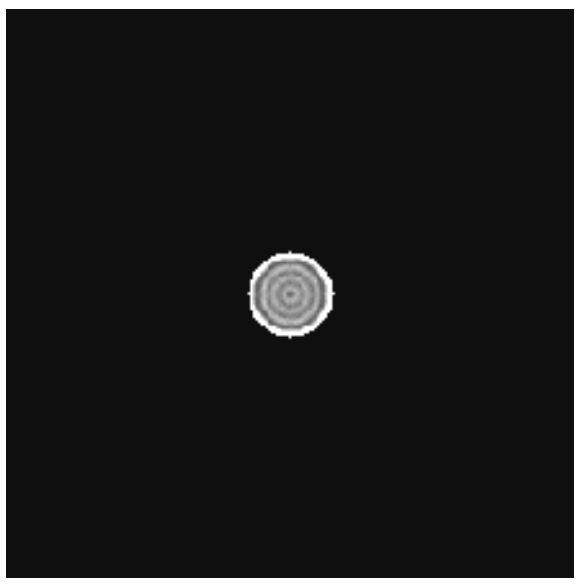

LP DFT-Sinclet spectrum ; N=256x256;SzLim=0.5; Blim=0.15

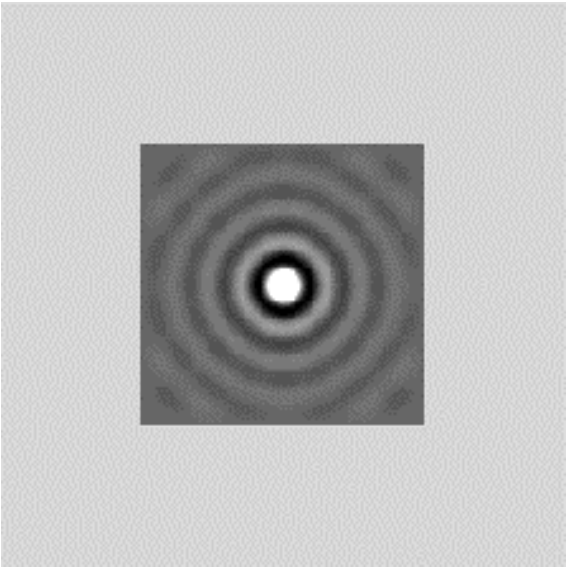

LP DFT-Sinclets: N=256x256;SzLim=0.5; Blim=0.15

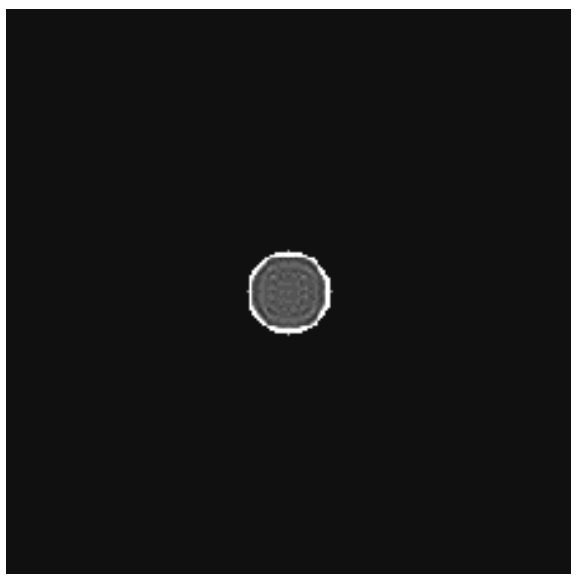

LP DFT-Sinclet spectrum; N=256x256;SzLim=0.5; Blim=0.15

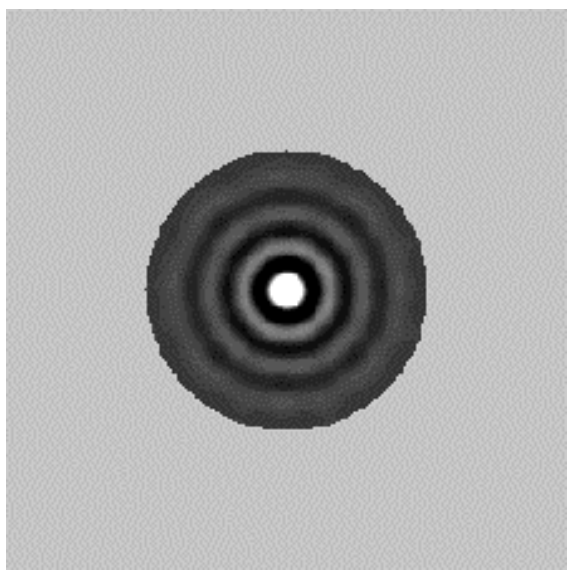

LP DCT-sinclets:N=256x256; SzLim=0.5; Blim=0.15

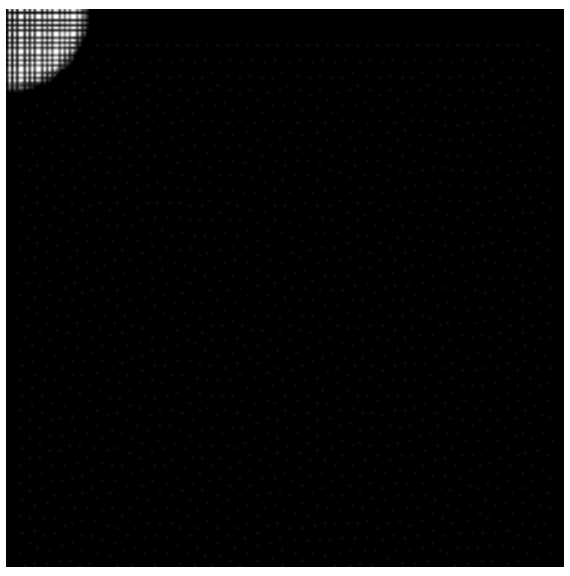

LP DCT-Sinclet spectrum:N=256x256; SzLim=0.5; Blim=0.15

Fig. 12. Examples of 2D LP sinclets (left column) and of their DFT and DCT spectra (right column). Size limitation (Slim) and band limitation (Blim) are given in fractions of the size of the corresponding domain.